\documentclass[aps,preprint]{revtex4}
\usepackage{graphics}
\def\XY{{\em XY} }
\def\KT{Kosterlitz-Thouless }
\def\2D{two-dimensional }
\def\3D{three-dimensional }
\def\k{\tilde{\kappa}}

\begin{document}

\title{Invaded cluster simulations of the \XY model in two and three 
dimensions }
\author {I. Dukovski}
\author {J. Machta}
\email{machta@physics.umass.edu}
\address{Department of Physics,
University of Massachusetts, Amherst, Massachusetts 01003-3720}
\author{L. V. Chayes}
\email{lchayes@math.ucla.edu}
\address{Department of Mathematics, University of California, Los
Angeles, California 90095-1555}
\date{\today}

\begin{abstract}
The invaded cluster algorithm is used to study the \XY model in two
and three dimensions up to sizes $2000^2$ and $120^3$ respectively.
A soft spin $O(2)$ model, in the same universality class as the \3D \XY model,
is also studied.
The static critical properties of the model and the dynamical properties
of the algorithm are reported.
The results are
$K_{c}=0.45412(2)$ for the \3D \XY model
and $\eta=0.037(2)$  for the \3D \XY universality class.
For the \2D \XY model the results are $K_{c}=1.120(1)$ and 
$\eta=0.251(5)$. The invaded
cluster algorithm  does not show any critical slowing for
the magnetization or critical temperature estimator for the \2D
or \3D \XY models.
\end{abstract}

\maketitle



\section{Introduction}

In this paper we present a Monte Carlo study of the \XY model using the invaded
cluster (IC) algorithm.  The purpose of this study is both to obtain high
precision results for the \XY model and to test the efficacy of the invaded
cluster method for phase transitions with continuous symmetry breaking.   The
\XY model is in the $O(2)$ universality class and in two dimensions 
the transition is
of the \KT type~\cite{KoTh}. Theoretical and computational studies of the
$O(2)$ critical point include high temperature series expansions~\cite{BuCo},
renormalization-group calculations~\cite{GuZi} and Monte Carlo (MC)
simulations~\cite{Janke90,GoHa93,HaTo99,CHPRV}. The $\lambda$ transition in $^4$He is
also in the $O(2)$ universality class and the specific heat exponent $\alpha$
for this system has been measured to very high precision~\cite{LiSw96,GoMu93,SwCh92}.

Recent Monte Carlo studies of the \XY transition use versions of the 
Wolff algorithm~\cite{Wolff} because  near the critical point they are much more
efficient than local algorithms. The Wolff algorithm is an example of a
cluster algorithm of the kind first introduced by Swendsen and
Wang~\cite{SwWa,NeBa99} for Ising-Potts models. The
central idea of cluster algorithms is to identify clusters
of sites by a bond percolation process correlated to the spin
configuration.  The
spins of each cluster are then independently flipped.  Cluster algorithms
can be
extremely efficient when the percolation process
that defines the clusters has a percolation threshold that coincides with
the phase transition of the spins. This situation holds for the original
Swendsen-Wang algorithm applied to Ising-Potts models and was later shown to
hold for a variety of other spin systems with discrete
symmetries~\cite{KaDo,ChMa97a,ChMa98a}.   Wolff~\cite{Wolff} showed 
how to extend
cluster algorithms to spin models with
$O(n)$-symmetry by an Ising embedding method.

Invaded cluster (IC) algorithms~\cite{MaCh95,MaCh96,MoMaCh} are 
cluster algorithms with the property that they find and simulate the
critical point automatically without {\em a priori} knowledge of the critical
temperature.  The critical temperature is a direct output of the
algorithm and the magnetic exponents can be obtained from finite-size
scaling of the cluster size distribution.  In addition to providing the
critical
temperature and magnetic exponent without having to invoke methods such as
histogram re-weighting, IC algorithms appear to have less critical slowing than
corresponding Swendsen-Wang or Wolff algorithms~\cite{MoMaCh}.  IC algorithms
can, in principle, be constructed whenever a conventional cluster algorithm is
available with the property that the bond percolation process has a percolation
threshold that coincides with the phase transition.  IC algorithms have
thus far
been applied to systems with discrete symmetry breaking including
Ising-Potts~\cite{MaCh95,MaCh96,LiGl,FrCaCo} models, Widom-Rowlinson
models~\cite{JoGoMaCh} and the fully frustrated Ising model~\cite{FrCaCo}.
In this paper we show how Wolff's embedding scheme can
be used to construct an efficient IC algorithm for simulating the $O(2)$
critical
point in three dimensions and the \KT point in two dimensions.

One of our objectives is to obtain a high precision value of the $\eta$
exponent for the $O(2)$ universality class.  Because $\eta$ is itself very
small, it has proved difficult to measure it with much precision.  One of the
difficulties is the presence of corrections to scaling.  Recently, Hasenbusch
  and T\"{o}r\"{o}k \cite{HaTo99,CHPRV}
proposed a method to minimize this difficulty by considering a soft spin $O(2)$
model with a parameter controlling the variance in the length of the spin
vectors.  By adjusting this parameter they  minimize corrections to scaling
and improve their estimate of $\eta$.  Below we apply a modified 
version of the IC
algorithm to the soft spin model.

In the next Section we describe the IC algorithm for the \XY model.  In Sec.\
\ref{sec:results} we give results for the critical temperature, magnetic
exponents and dynamic properties of the \XY model in two and three 
dimensions. In Sec.\
\ref{sec:soft} we describe the algorithm and present results for the 
soft spin model.
Conclusions are presented in Sec.\ \ref{sec:conclusions}.

\section{Invaded cluster algorithm for the \XY model}

The algorithms for the {\em XY} model used in this paper are obtained by
combining the invaded cluster method and Wolff's embedding scheme for
continuous spin models.  The \XY model is defined by the Hamiltonian:
\begin{equation}
\beta H=-K\sum_{<i,j>}{\vec{s}}_{i}\cdot {\vec{s}}_{j}
\end{equation}
where ${\vec{s}}_{i}$ is a two dimensional unit vector and the summation
is over all nearest neighbor bonds on the lattice, here either the square or
cubic lattice with periodic boundary conditions.

We begin by describing a version of Wolff's
algorithm for the {\em XY} model.  On each Monte Carlo step we choose 
a two-dimensional
unit vector
$\vec{r}$.   For each bond $(i,j)$ of the lattice, the bond is called {\em
satisfied} if both spins lie on the same side of the line 
perpendicular to the unit vector,
that is,
\begin{equation} ({\vec s}_{i}\cdot {\vec r})({\vec s}_{j}\cdot {\vec r})>0.
\label{eq:sat}
\end{equation}
Satisfied bonds are then {\em occupied} with probability
\begin{equation}
\label{eqn:wprob}
P({\vec s}_{i},{\vec s}_{j})=1-\exp[-2K({\vec r}\cdot
{\vec s}_{i})({\vec r}\cdot{\vec s}_{j})].
\end{equation}
One way to implement the occupation of bonds with this probability is to
independently assign random numbers $u_{i,j}$, uniformly chosen from the
interval
$[0,1)$ to every bond of the lattice and then to occupy the satisfied bonds if
$u_{i,j}<P({\vec s}_{i},{\vec s}_{j})$. An equivalent approach, which
provides a
useful link to the IC methodology, is to define
$\kappa_{i,j}$ from $u_{i,j}$ by substituting $u_{i,j}$ for $P({\vec
s}_{i},{\vec
s}_{j})$ and $\kappa_{i,j}$ for $K$ in Eq.\
(\ref{eqn:wprob}) and then solving for $\kappa_{i,j}$,
\begin{equation}\kappa_{i,j}=-\log(1-u_{i,j})/(2({\vec r}\cdot
{\vec s}_{i})({\vec r}
\cdot{\vec s}_{j})). \label{eq:K}
\end{equation}
Satisfied bonds with $\kappa_{i,j} < K$ are occupied. The occupied bonds define
a set of connected clusters. Single sites with no occupied bonds are considered
clusters so that every lattice site is uniquely a member of some cluster.

Once the clusters are identified each cluster is {\em
flipped} with probability $1/2$.  A cluster is flipped by reflecting every spin
in the cluster through the line perpendicular to ${\vec r}$,
\begin{equation}
\label{eq:flip}
{\vec s}_{i} \rightarrow R({\vec r}){\vec s}_{i}
\end{equation}
where
\begin{equation}
\label{eq:reflect}
R({\vec r}){\vec s}_{i}={\vec s}_{i}-2({\vec
s}_{i}\cdot {\vec r}){\vec  r}.
\end{equation}
Flipping clusters with probability one half yields a new spin
configurations and
completes one MC step. It is straightforward to show that the algorithm is
ergodic and satisfies detailed balance.

Wolff's original paper~\cite{Wolff} introduces two innovations.  The first
is to grow and flip only a single cluster in each Monte Carlo step.  The
second is the generalization of cluster methods to spin models with
continuous symmetries.  Only the generalization to continuous
symmetries is used here and combined with the invaded cluster
methodology.  In fact, Wolff's single cluster method is not compatible with the
invaded cluster methodology.  When we speak of the Wolff algorithm in the
following we mean the algorithm described above for which all clusters are
defined and flipped in a single MC step.

Cluster algorithms can generally be viewed as a sequence of bond moves
and spin moves.  During the bond move a configuration of occupied bonds and
the associated set of clusters are generated from the spin
configuration.  This is done by occupying satisfied bonds with a
probability that depends on the simulation temperature.  During the spin move a
new spin configuration is obtained  by randomly flipping clusters.  The bond
configurations can be viewed as a correlated percolation model.  For a cluster
algorithm to be efficient, the correlated percolation model  should have a
percolation transition that coincides with the critical point of the spin
model.  If this holds then clusters of all sizes are flipped during a single
MC step and changes to the spin configuration occur on all length scales.  For
Ising-Potts models the correlated percolation model associated with the
Swendsen-Wang algorithm is the Fortuin-Kastelyn random
cluster model and the equivalence of the percolation threshold and the
critical point is well-understood
\cite{CoKl,ACCN}.  The percolation properties of the bonds defined by the
Wolff algorithm for the \XY model have recently been
investigated~\cite{Chayes98}.  The conclusion is the same as for the
Ising-Potts
models: the critical point for the \3D {\em XY} model and Kosterlitz-Thouless
point for the \2D {\em XY} model coincide with the percolation 
threshold for the
bonds defined by the Wolff algorithm.

The invaded cluster methodology relies on the equivalence of the phase
transition in the spin model and the percolation transition of the occupied
bonds to find and simulate the phase transition without prior knowledge of
the critical coupling. Like other cluster algorithms, a full MC step
consists of a bond move followed by a spin move.  The spin move is the
same as for standard cluster algorithms but the bond moves differs.  In the IC
bond move, satisfied bonds are occupied one at a time in random order until a
signature of percolation is first observed.  The set of occupied bonds obtained
in this way defines bond clusters and these are flipped in the usual way. The
signature of percolation is incorporated in a stopping condition that is tested
after each new bond is occupied.  In this paper we use a topological stopping
condition, which requires that at least one cluster wraps around the 
lattice in at
least one direction.

For the most part, the IC algorithm closely parallels the Wolff algorithm.
First a unit vector is randomly chosen and satisfied bonds are determined with
respect
to this unit vector as described in the paragraph including Eq.\ 
(\ref{eq:sat}).
Then
uniform random numbers,
$u_{i,j}$ are assigned to each bond and non-uniform random numbers,
$\kappa_{i,j}$ are obtained from them according to Eq.\ (\ref{eq:K}).  The next
step of the IC algorithm differs from the Wolff algorithm.  Satisfied bonds are
occupied {\em one at a time} according to the ordering defined by
$\kappa_{i,j}$
with the satisfied bond having the smallest value of $\kappa$ occupied first.
After each bond is occupied, the set of clusters is updated and the stopping
condition is checked.  If no cluster wraps around the lattice the satisfied
bond
with the next largest
$\kappa$ is occupied but if some cluster wraps around the lattice in some
direction the bond move is completed and no further bonds are occupied.  The
largest value of $\kappa$ chosen during the bond move is called $\k$.  The spin
move is identical to the Wolff algorithm: with probability
$1/2$, each cluster is reflected through the line perpendicular to 
${\vec r}$ according to
Eqs.\ (\ref{eq:flip}) and (\ref{eq:reflect}).

The IC algorithm simulates the critical point and $\k$ is an estimator of
the critical coupling.  To see why this is the case, consider a large
system with
a spin configuration that is typical of the critical point.  The correlated
percolation threshold for the occupied bonds is related to critical coupling
according to Eq.\ (\ref{eqn:wprob}). That is, if satisfied bonds are occupied
with probability
$P_c({\vec s}_{i},{\vec s}_{j})=1-\exp[-2K_c({\vec r}\cdot  {\vec s}_{i})({\vec
r}\cdot{\vec s}_{j})]$, the occupied bonds will just
percolate~\cite{Chayes98}.
Occupying bonds with probability $P_c({\vec s}_{i},{\vec s}_{j})$ is the
same as
occupying all satisfied bonds one at a time in ascending order of $\kappa$ and
stopping at the largest $\kappa$ such that $\kappa \leq K_c$. The IC
algorithm works in the same way except that satisfied bonds are added until a
cluster wraps around the system.  For a large system, this event will occur
with
$\k$ nearly equal to $K_c$.  Thus, if the system
is at criticality, a single step of the IC algorithm is almost identical to
a single step of the Wolff algorithm at $K_c$ and $\k \approx K_c$.
So long as the fluctuations in $\k$ become small as the system size
increase, the IC algorithm and the Wolff algorithm at $K_c$ will sample
the same state in the thermodynamic limit.  However, in finite volume, the
invaded cluster algorithm defines an ensemble that is different than the
canonical ensemble and has different finite-size scaling properties.

So far we have assumed the spin system is already at the critical point.
Suppose that the spin configuration is typical of the low temperature
phase, then
there will be more satisfied bonds than at criticality and a smaller fraction
will be needed to form a spanning cluster so that
$\k < K_c$.  The invaded cluster MC step thus corresponds to a
Wolff MC step at temperature $T=J/\k > T_c$ so that the system is warmed.  A
similar argument shows that if the system starts in the high temperature phase,
it is cooled by the IC algorithm so that there is negative feedback mechanism
that forces the system to the critical point independent of the starting
configuration. A detailed  discussion of these arguments in the context of
Ising-Potts models is given in Ref.\ \cite{MaCh96}.

For the 2D \XY model, the IC algorithm described above does not 
perform well.  The problem
is similar to a problem that occurs in the 2D Ising model. 
Specifically, at criticality,
the satisfied bonds themselves just percolate. This means that in a 
significant fraction of
spin configurations, spanning is not possible.  It also means that 
the distribution of the
temperature estimator is broad because of spin configurations for 
which spanning is just
barely possible. The problem is even worse for the
the \XY model.  Since the unit vector
${\vec r}$ is randomized from one step to the next, spanning on one 
Monte Carlo step does
not guarantee spanning on the next Monte Carlo step as it does in the 
Ising case.  As a
result, a small fraction of the time, no spanning cluster can be 
found for the chosen unit
vector
${\vec r}$.  Furthermore, the distribution of the temperature 
estimator is very broad
because some spin configurations and unit vectors just barely allow 
spanning yielding a
very small value of the temperature estimator.

These problems can be alleviated by taking
advantage of a second independent set of satisfied bonds. Let's 
refer to the bonds
that are satisfied with respect to the definition of Eq.\ 
(\ref{eq:sat}) as {\em red}
satisfied bonds.  Let ${\vec b}$ be a unit vector perpendicular to 
${\vec r}$ and
define bond $(i,j)$ as {\em blue} satisfied if
\begin{equation} ({\vec s}_{i}\cdot {\vec b})({\vec s}_{j}\cdot {\vec b})>0.
\label{eq:bsat}
\end{equation}
To increase the probability that we obtain a spanning cluster, we can 
occupy both red and
blue satisfied bonds.  Two values of $\kappa$, one with respect to
${\vec r}$ and the other with respect to ${\vec b}$, are assigned to 
each bond using Eq.\
(\ref{eq:K})  and a single value of $u$.  The entire set of 
$\kappa$'s, for both red
and blue satisfied bonds, is ordered.  Bond are either red or blue 
occupied in the order
prescribed by the $\kappa$'s and sets of red and blue clusters are 
identified and updated
after each new bond is occupied.  The first cluster to span, either 
red or blue, stops the
process of occupying bonds.  During the spin move, both sets of 
clusters are flipped with
probability $1/2$ according to Eqs.\ (\ref{eq:flip}) and 
(\ref{eq:reflect}) with ${\vec r}$
replaced by ${\vec b}$ for the blue clusters.  One way of looking at 
this algorithm is that
it utilizes two independent embeddings of the Ising model in the XY model, one
embedding relative to ${\vec r}$ and the other relative to ${\vec 
b}$.  The two embedding
method was always able to find a spanning cluster and it produces a less
broad distribution for the temperature estimator.  Nonetheless, the 
distribution of
$\k$ is still very broad and we find better statistics by averaging 
$1/\k$. All results
reported for the two-dimensional \XY model use the two-embedding method.

\section{Results}
\label{sec:results}

The algorithm was implemented on systems with maximum linear size $L=120$
for three dimensions and $L=2000$ for two dimensions. Each run 
consisted of 160000 MC steps for three dimensions and 10000 MC 
steps for two dimensions .   We collected
statistics for the estimator of the critical coupling, $\k$ and
the size of the spanning cluster, $M$. For the \2D model we obtained 
the estimator
of critical coupling as inverse of the estimator of the critical temperature.
The average critical coupling and its error
bars were obtained using the blocking
method~\cite{NeBa99} with 100 blocks of 100 data points each for two dimensions
and 160 blocks of 1000 data points each for three dimensions.
The average size of the spanning cluster and its error
bars was obtained using the bootstrap
method~\cite{NeBa99}.
The reported error bars are one standard deviation.
The simulation time was $10^{-5}$ CPU seconds  per spin per MC sweep on a
$450$MHz Pentium III machine running Linux.

\subsection{Critical temperature of the \3D \XY model}

\begin{figure}
\includegraphics{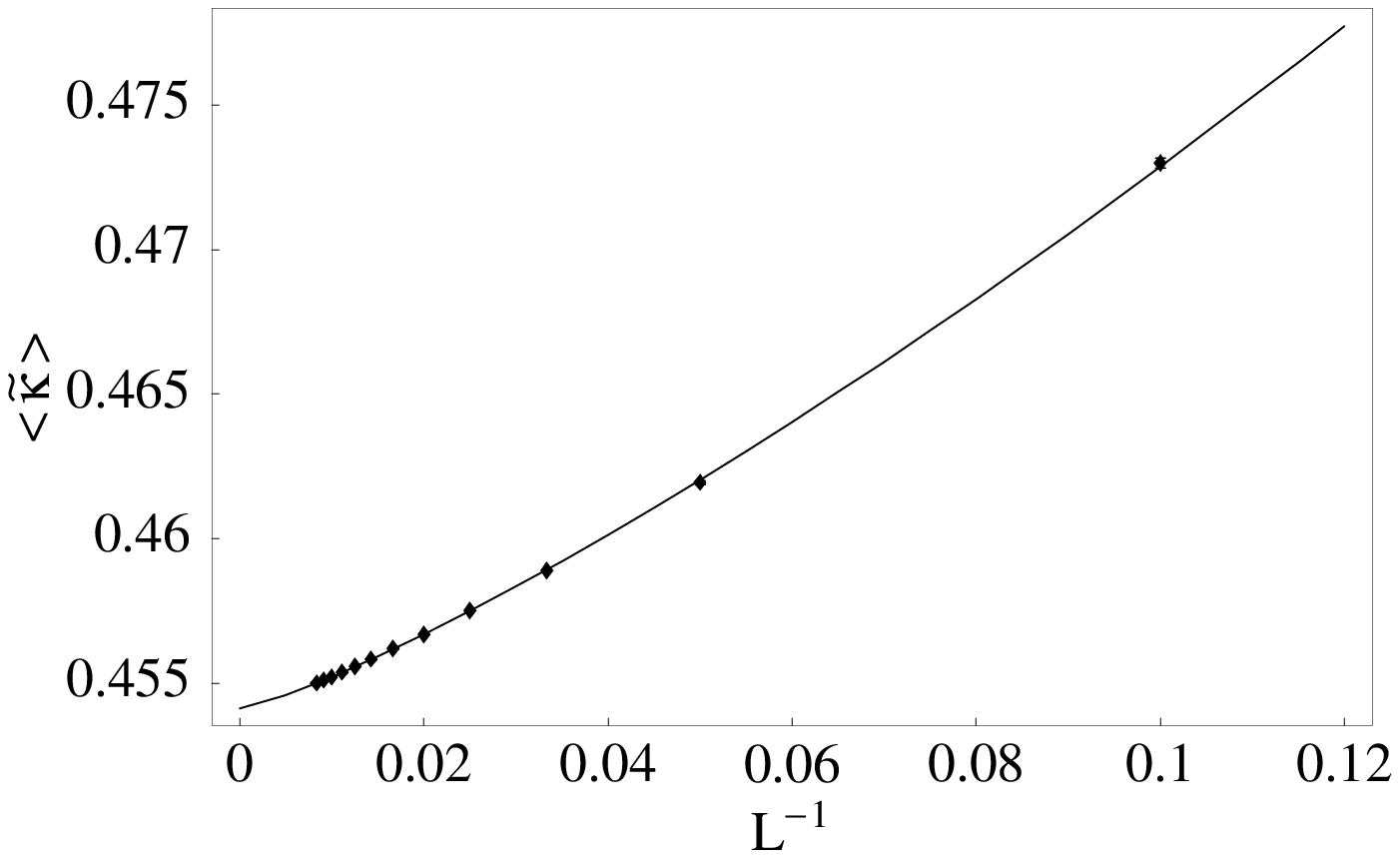}
\caption{Critical coupling $\langle \k \rangle$ vs.\ $1/L$ for the
three-dimensional
\XY model.  The solid line is a fit to
the data as described in the text. }
\label{fig:Kc3d}
\end{figure}

Figure~\ref{fig:Kc3d} shows the average value of $\k$ for the \3D\ \XY model
versus the inverse of the linear size of the system $1/L$.
We fit the data for $L \geq 10$ to a function of the form:
\begin{equation}
\langle \k(L) \rangle = \frac{K_c}{1+a L^{-p}}
\end{equation}
where $K_c$, $a$ and $p$ are parameters of the fit.
  The results of the fit are
$K_c=0.45412(2)$, $a=-0.64(2)$ and $p=1.211(9)$ with $\chi^2=6.23$, 
$\chi^2/d.o.f.=0.69$
and the confidence level is $Q=0.72$.
The value for the critical coupling $K_c$ compares well with values used 
in the literature as shown in Table~\ref{tab:3d}. 
Note that $p \neq 1/\nu \approx 1.5$ as would be
expected from naive finite size scaling arguments.  As is the case 
for Ising-Potts
systems, the IC ensemble for the \XY model does not have the same 
finite size scaling
properties as the canonical ensemble.

\begin{figure}
\includegraphics{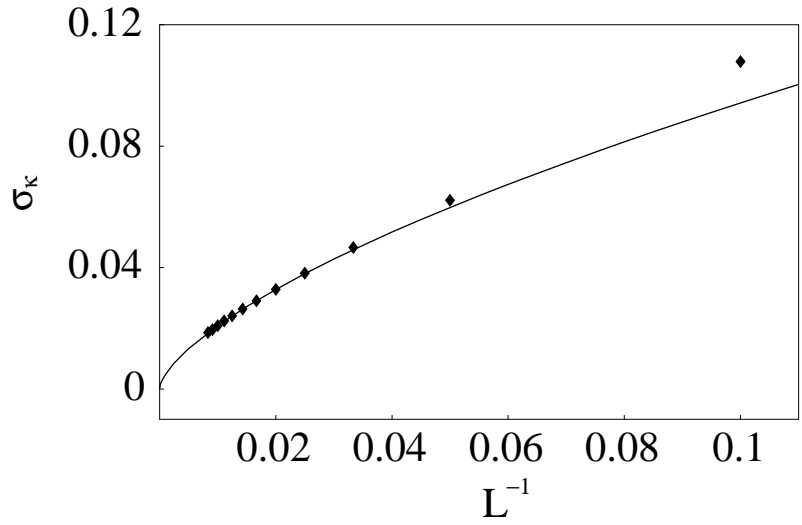}
\caption{Standard deviation of the critical coupling $\sigma_{\k}$
  vs.\ $1/L$  for \3D XY model. The solid line is a fit to
the data as described in the text.}
\label{fig:sig3d}
\end{figure}

The validity of the IC method is justified by the fact that the width of the
distribution of $\k$ decreases as $L$ increases.
Figure~\ref{fig:sig3d} shows the standard  deviation $\sigma_{\k}$ of 
$\k$ as a function of
$1/L$ and suggests that
\begin{equation}
\lim_{L\rightarrow\infty}\sigma_{\k}= 0
\end{equation}
and therefore we expect a sharp distribution of $\k$ for $L=\infty$. 
The solid curve in
Fig.~\ref{fig:sig3d} is the result of a fit to the data for $L \geq 
50$ to the functional
form $\sigma_{\k} = a + b L^{-q}$ yielding $a=0.0002(10)$, $b=0.43(3)$ and $q=0.66(3)$.

\subsection{Kosterlitz-Thouless temperature}

\begin{figure}
\includegraphics{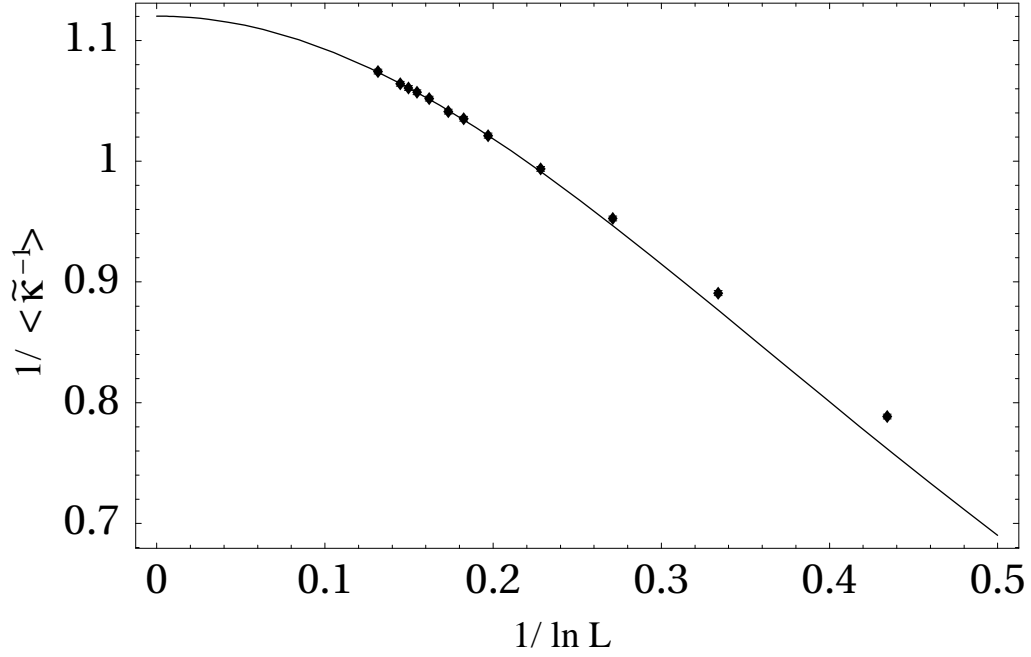}
\caption{Critical coupling 1/$\langle \k^{-1} \rangle$ vs.\ 
$1/\ln(L)$ for \2D \XY
model. The solid line is a fit  to the data as described in the text.}
\label{fig:kl2d}
\end{figure}

Figure~\ref{fig:kl2d} shows $\langle 1/\k(L) \rangle^{-1}$ as a 
function of $1/\log (L)$
for the \2D \XY model. We fit the data for $L \geq 160$ to the function,
\begin{equation}
\langle 1/\k(L) \rangle^{-1} =\frac{K_c}{1+a(\log(L))^{-2}}
\end{equation}
where $K_c$ and $a$ are parameters of the fit.
This functional form was motivated by combining the usual finite size 
scaling assumption
that $\xi=L$ with the Kosterlitz-Thouless  expression for the 
critical behavior of the
correlation length
$\xi$
\cite{Ko74},
\begin{equation}
\xi_{\infty}(K)\sim e^{bt^{-\nu}}
\end{equation}
where $t$ is the reduced temperature and  $\nu=1/2$.
For this fit we found $K_c=1.120(1)$ and $a=2.49(4)$ with
$\chi^2=3.9$, $\chi^2/d.o.f.=0.65$ and $Q=0.68$.
The choice of $L \geq 80$ was a compromise between keeping the error 
in $K_c$ small
and the confidence level $Q$ large. The result for $K_c$ is in good agreement
with some of the recent results from the literature as shown in Table~\ref{tab:2d}.
 Although the fit is good, the 
result for $K_c$ should
be viewed with caution because it is based on finite size scaling 
assumptions that do not
necessarily hold for the IC ensemble.

\begin{figure}
\includegraphics{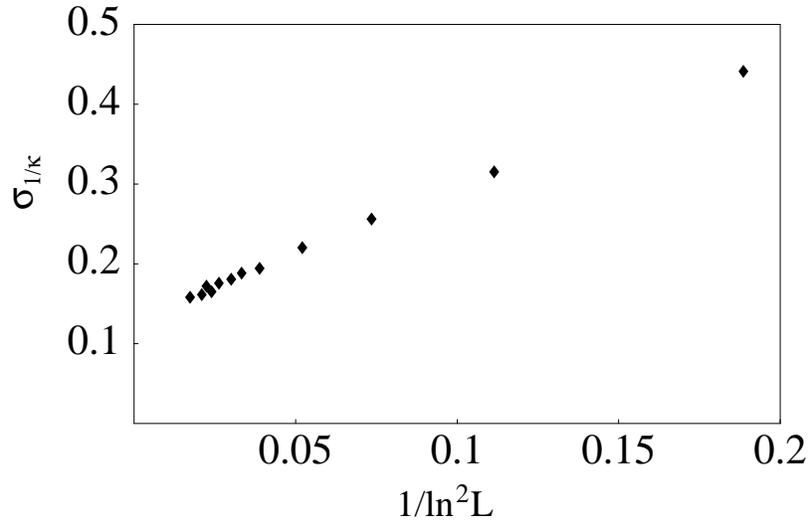}
\caption{Standard deviation of the critical coupling $\sigma_{1/\k}$
  vs.\ $1/\ln^2(L)$  for \2D \XY model. }
\label{fig:sig2d}
\end{figure}

Figure~\ref{fig:sig2d} shows the standard
deviation $\sigma_{1/\k}$ of $1/\k$ as a function of $1/\log^2(L)$. Unlike 
the situation in
three-dimensions, it is not clear that $\sigma_{1/\k}$ vanishes as $L
\rightarrow
\infty$. 
If the finite size scaling behavior of the IC ensemble is of the 
``essential singularity''
type observed in the canonical ensemble then a reasonable hypothesis is that
$\sigma_{1/\k} \sim (\log L)^{-q}$.
 A naive extrapolation suggests $\sigma_{1/\k}$ approaches a 
finite value near
$0.11$ but a slowly decreasing function cannot be ruled out. In the 
previous section we
discussed reasons for the broad distribution of the temperature 
estimator.  
Simulations with
$L \gg 2000$ would be needed to determine whether or not
$\sigma_{1/\k}$ vanishes as $L\rightarrow \infty$.

\subsection{Magnetic Exponents}

The average mass $M$ of the spanning cluster is proportional to the 
magnetization of the
system so that the magnetic exponents can be obtained from the 
fractal dimension of the
spanning cluster, defined by $M \sim L^D$.  The
critical exponent $\eta$ is related to $D$ via,
\begin{equation}
\eta=2+d-2D.
\end{equation}

\begin{figure}

\includegraphics{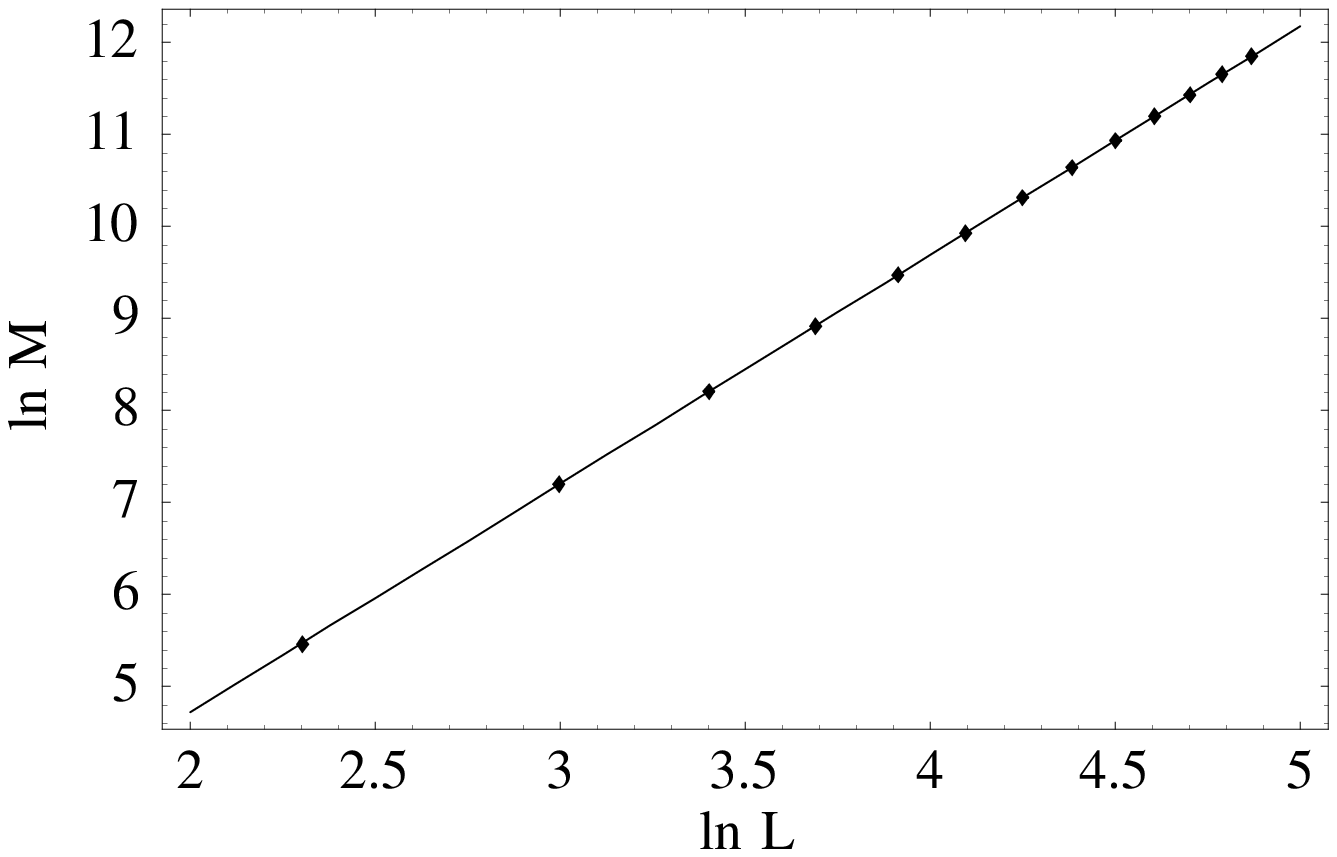}
\caption{Mass of the spanning cluster $\ln M$ vs.\ $\ln L$ for the \3D \XY model.
The solid line is a linear fit.}
\label{fig:ML3d}
\end{figure}

\begin{figure}
\includegraphics{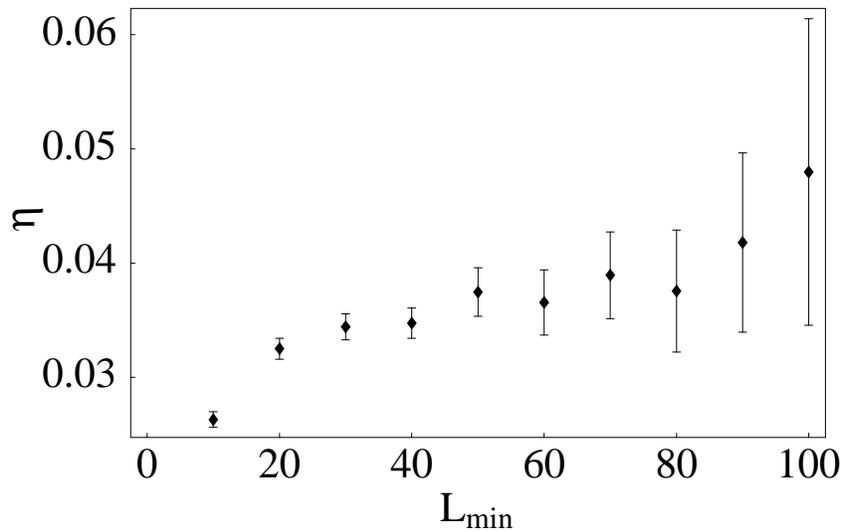}
\caption{The critical exponent $\eta$ for the \3D \XY model vs.\ minimum system size $L_{min}$ 
included in the fit of the data.
The maximum system size in the fit is $L_{max}=120$.}
\label{fig:trend}
\end{figure}

Figure~\ref{fig:ML3d} shows a log-log plot of $M$ vs.\  $L$ for the 
three-dimensional \XY
model. A linear fit yields $\eta(3D)=0.037(2)$ with $\chi^2=2.1$,
$\chi^2/d.o.f.=0.35$ and the confidence level $Q=0.9$. The smallest 
value of $L$ included
in the fit is  $L_{min}=50$.
Figure~\ref{fig:trend} shows values of $\eta$ obtained from fits with 
smallest linear size $L_{min}$. The value of $\eta$ has an upward trend
up to $L_{min}=40$ and then is constant for larger values of $L_{min}$.
The value of $L_{min}$ for the reported results was  chosen such that
the statistical error is minimal for a reasonably large $Q$ and the value of $\eta$
is well into the region of constant values. The plot on Fig.~\ref{fig:trend} also shows
an upward trend for the two last values of $L_{min}$. 
Table~\ref{tab:3d} shows some recent results for the \3D \XY model.
Our result for $\eta$ agrees with recently published values.

\begin{figure}
\includegraphics{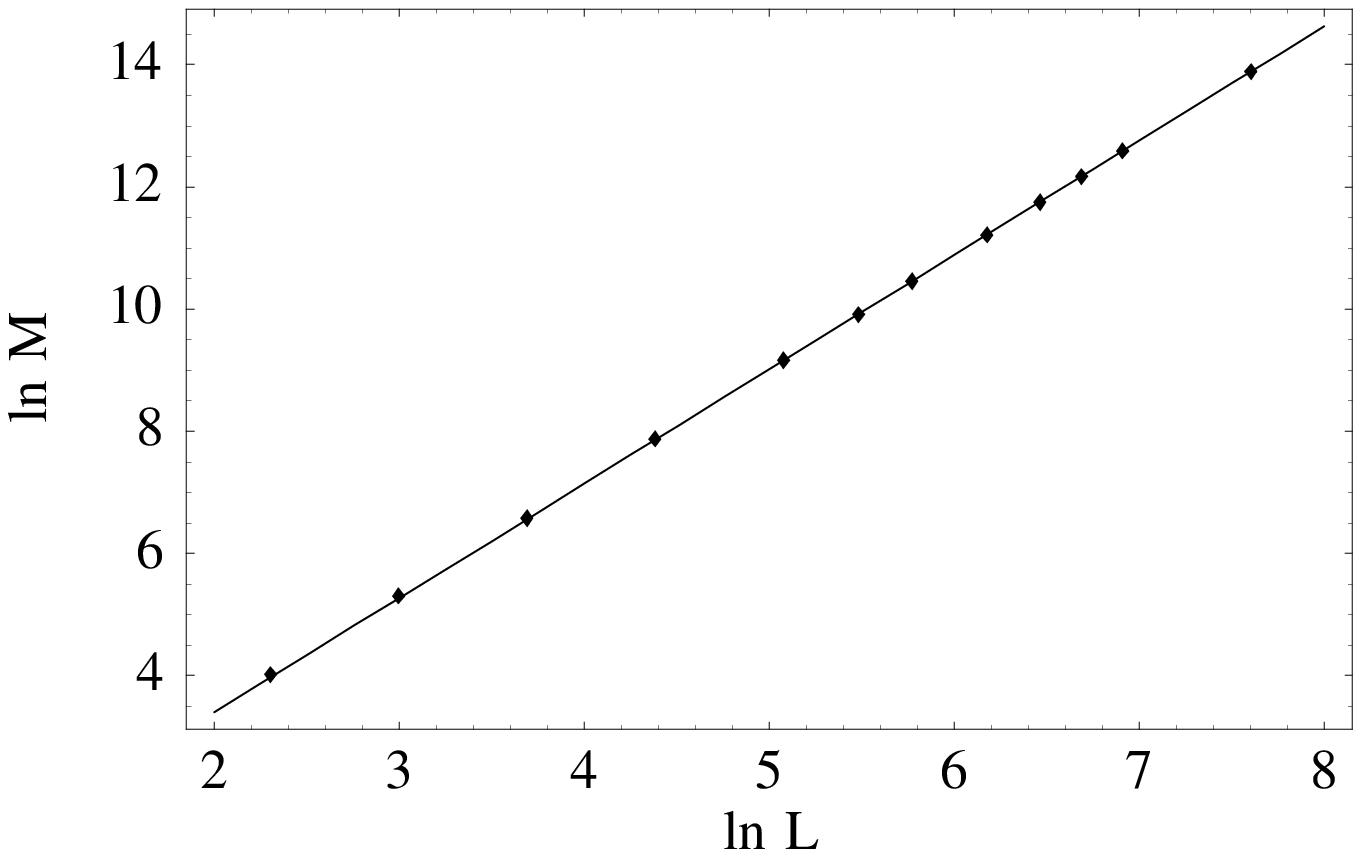}
\caption{Mass of the spanning cluster $\ln M$ vs.\ $\ln L$ for \2D \XY 
model. The solid
line  is a linear fit.  }
\label{fig:ML2d}
\end{figure}

Figure~\ref{fig:ML2d} shows a log-log plot of $M$ vs.\  $L$ for the 
two-dimensional \XY
model. A linear fit to the data yields $\eta(2D)=0.251(5)$ which is in
good agreement with the theoretical value for the Kosterlitz-Thouless 
transition
$\eta_{KT}=0.25$. The smallest value of $L$ included in the fit is
$L_{min}=480$ and $\chi^2=2.84$, $\chi^2/d.o.f=0.94$ and
$Q=0.41$. The value of $L_{min}$ was chosen such that
the statistical error is minimal for a reasonably large $Q$.

\subsection {Dynamics of the invaded cluster algorithm}
\label{sec:dyn}
The {\em autocorrelation function} for a time dependent
variable  $A(t)$ is defined as:
\begin{equation}
\Gamma_{A}(t)=\frac{<(A(0)-<A>)(A(t)-<A>)>}{<(A(0))-<A>)^2>}
\end{equation}
The  {\em integrated autocorrelation time} is defined by
\begin{equation}\tau_{A}=\frac{1}{2}+\lim_{w \rightarrow
\infty}\sum_{t=1}^{w}\Gamma_{A}(t).
\label{eq:tau}
\end{equation}
The integrated autocorrelation time for $A$
is interpreted as the time needed to obtain
statistically independent measurements of $A$.  As a result, the statistical
error in measuring $A$ in a MC simulation is proportional to
$(\tau_A/N)^{1/2}$ where $N$ is the number of MC steps.

\begin{figure}
\includegraphics{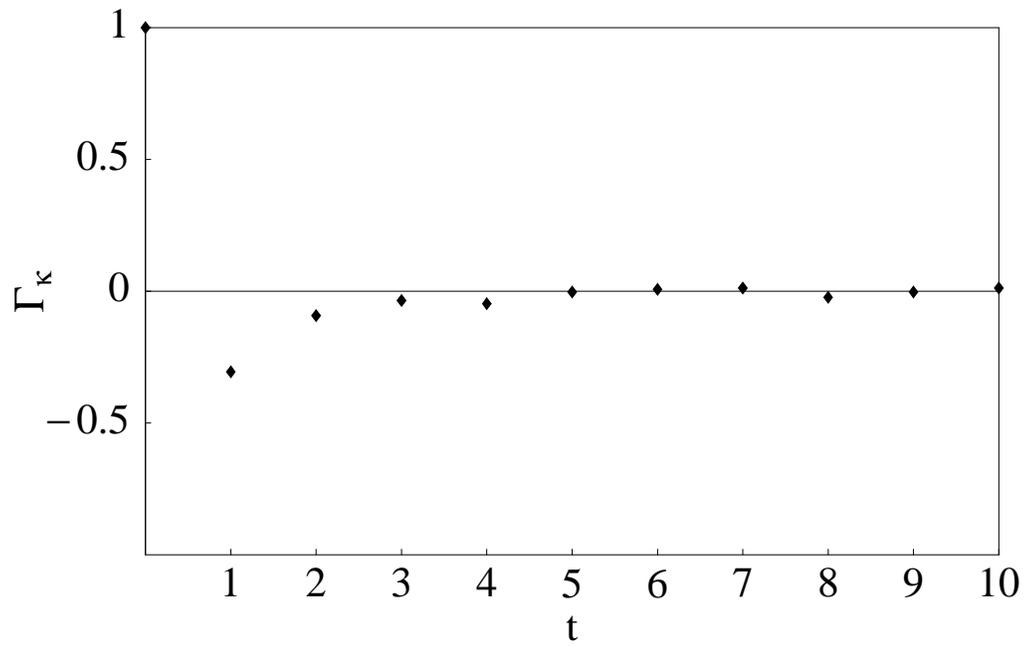}
\caption{The autocorrelation function of the critical coupling $\Gamma_{\k}$
vs.\ Monte Carlo time for the \3D \XY model. The 
system size is L=100.}
\label{fig:correl3d}
\end{figure}

\begin{figure}
\includegraphics{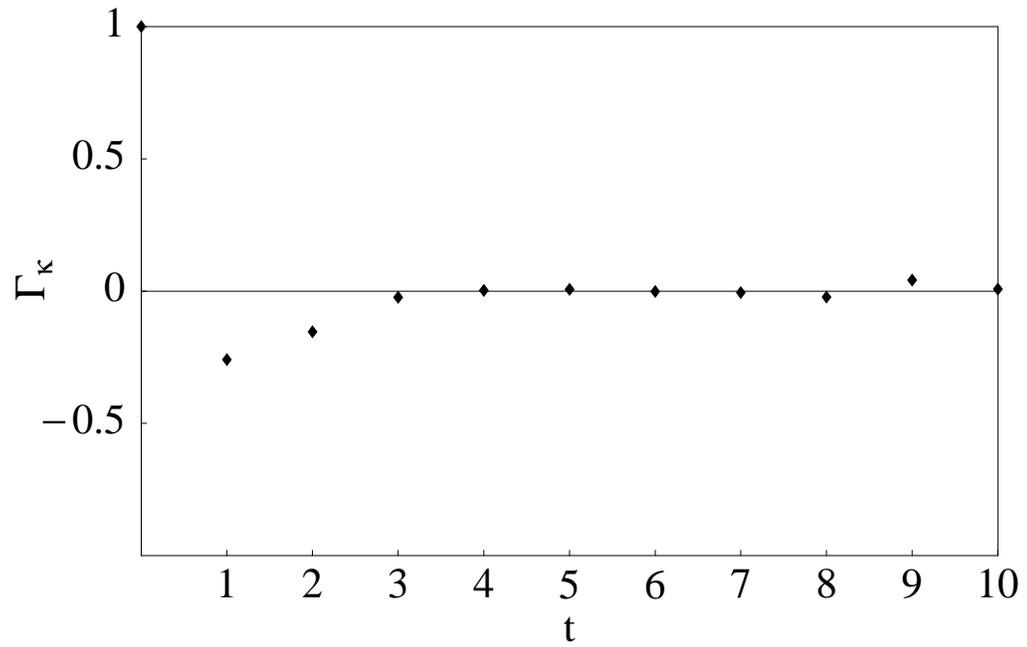}
\caption{
The autocorrelation function for the \KT temperature estimator 
$\Gamma_{\k}$ vs.\ Monte Carlo time for the \2D \XY model. The 
system size is L=1000.}
\label{fig:correl2d}
\end{figure}

We measured the autocorrelation functions and the
corresponding integrated autocorrelation times for the magnetization $M$ and
the  critical coupling estimator $\k$.  When calculating the integrated
autocorrelation time it is necessary to choose a
finite cut-off  for $w$ in the  sum in Eq.\ (\ref{eq:tau}). We used $w=100$, a
value much longer than the integrated autocorrelation times obtained below.
Figure~\ref{fig:correl3d} shows the autocorrelation function versus time for the
inverse critical  temperature of the \3D \XY model. Figure~\ref{fig:correl2d}
shows the same plot for the \2D case.  Note the negative  overshoot of the
correlation functions for time of one MC step. This is due to the  presence
of a
negative feedback mechanism in the IC algorithm. Consecutive 
temperature estimators
for the spin
configurations
are  anti-correlated.

Table~\ref{tab:tau3d} shows the integrated autocorrelation times  for 
the mass of
spanning cluster $\tau_M$ and the inverse temperature $\tau_{K_c}$ as 
a function of the
linear size of the system $L$ in the three-dimensional
\XY model. Table~\ref{tab:tau2d} shows the same data for the 
two-dimensional case. These
data show that the autocorrelation times for $M$ and $\k$  are 
independent of the  the
system size.  This behavior was also observed for IC dynamics for Ising-Potts
models~\cite{MaCh95,ChMaTaCh,MoMaCh} and gives the impression that 
the IC algorithm
does not have any critical slowing down.  However, as shown by 
Moriarty, Machta and
Chayes\cite{MoMaCh} for Ising-Potts models, observables defined on 
scales intermediate
between the lattice spacing and system size have relaxation times that 
diverge with system
size so that the dynamic exponent for the algorithm is greater than 
zero. Nonetheless,
the small values of
$\tau_{\k}$ and $\tau_M$ mean that relatively few MC steps are needed 
to obtain good
statistics for $\k$ and $M$.

\section{The soft spin $O(2)$ model}
\label{sec:soft}
The values of  $\eta$ for the \3D \XY universality class vary
significantly in the literature.
Table\ \ref{tab:3d} shows some recent results for $\eta$.
Possible reason this discrepancy in the literature 
is the presence of finite size 
scaling correction
in the fit of $M$ vs.\ $L$. Hasenbusch and 
T\"{o}r\"{o}k~\cite{HaTo99} and Campostrini \emph{at. al.}\cite{CHPRV}
 studied a
$\phi^4$ \textit{soft spin} model defined by the Hamiltonian:
\begin{equation}
H/T=-J/T\sum_{<i,j>}{\vec\phi}_{i}\cdot{\vec{\phi}}_{j}+
\sum_i{\phi_i}^2+
\lambda\sum_i({\phi_i}^2-1)^{2}
\end{equation}
where $T$ is the temperature, $J$ interaction constant and $\lambda$ 
a ``softness"
parameter. The vectors $\vec{\phi}$ can have any value of the modulus
$|\vec{\phi}|$. The \XY model is obtained as a special case for 
$\lambda=\infty$.
This model is in the same universality class as the \3D \XY model.
Hasenbusch and T\"{o}r\"{o}k showed that finite size
corrections in the canonical ensemble are minimized near $\lambda=2.0$.

We implemented the IC algorithm for the soft spin  model. Following Refs.\
\cite{BrTa} and \cite{HaTo99} we used the IC algorithm to update the 
orientation of the
vectors
$\vec{\phi}$ and the Metropolis algorithm to update the modulus. The 
IC part of the
combined algorithm is described  in Sec.~2. The only difference is 
that the modulus of
the spins does not necessarily have the value  of
$|\vec{\phi}|=1$. Since the IC algorithm performs only reflections,
the modulus of the spins
remains the same after the IC sweep.
After every IC step, the algorithm performs an update
of both the modulus and the orientation of the spins using the 
Metropolis algorithm as
follows. New values of the two spin components are proposed:
\[{\phi_x}'=\phi_x-2(p_x-0.5)\]
\[{\phi_y}'=\phi_y-2(p_y-0.5)\]
where $p_x$ and $p_y$ are random numbers from the interval $[0,1)$.
The update is accepted with the probability \[P=\min[1,\exp(H/T-H'/T)]\]
where $T$ is obtained from the output of the previous IC step.
The temperature will therefore be different  
for every Metropolis update. 
However, as is the case for the \3D \XY model, these temperature fluctuations 
tend toward zero as $L\rightarrow \infty$.
In the limit of very
large system size the Metropolis subroutine updates the system at a fixed 
temperature. The Metropolis subroutine preserves the negative feedback mechanism
of the IC algorithm. If the temperature of the system is low, the Metropolis algorithm
will tend to make the modulus of the spins larger. With larger spin modulus the
IC algorithm will give a higher temperature as an outcome. The same holds if we start with 
high temperature. The Metropolis algorithm will make the modulus of the spins smaller
and consequently the IC algorithm will lower the temperature.

\begin{figure}
\includegraphics{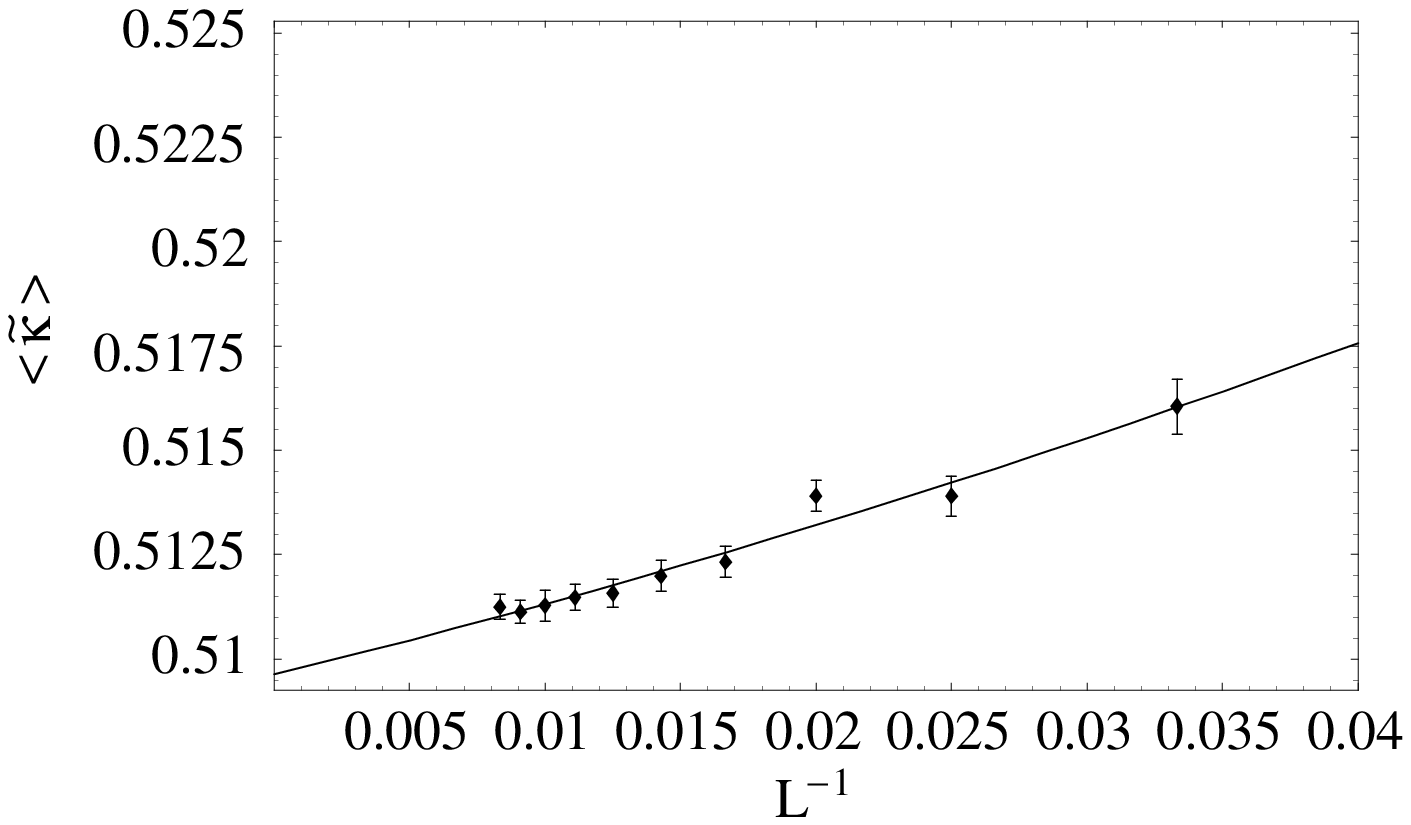}
\caption{Critical coupling $\langle \k \rangle$ vs.\ $1/L$ for the 
\3D soft spin model. The solid line is a fit
to the data as described in the text.
}
\label{fig:softK}
\end{figure}

\begin{figure}
\centering
\includegraphics{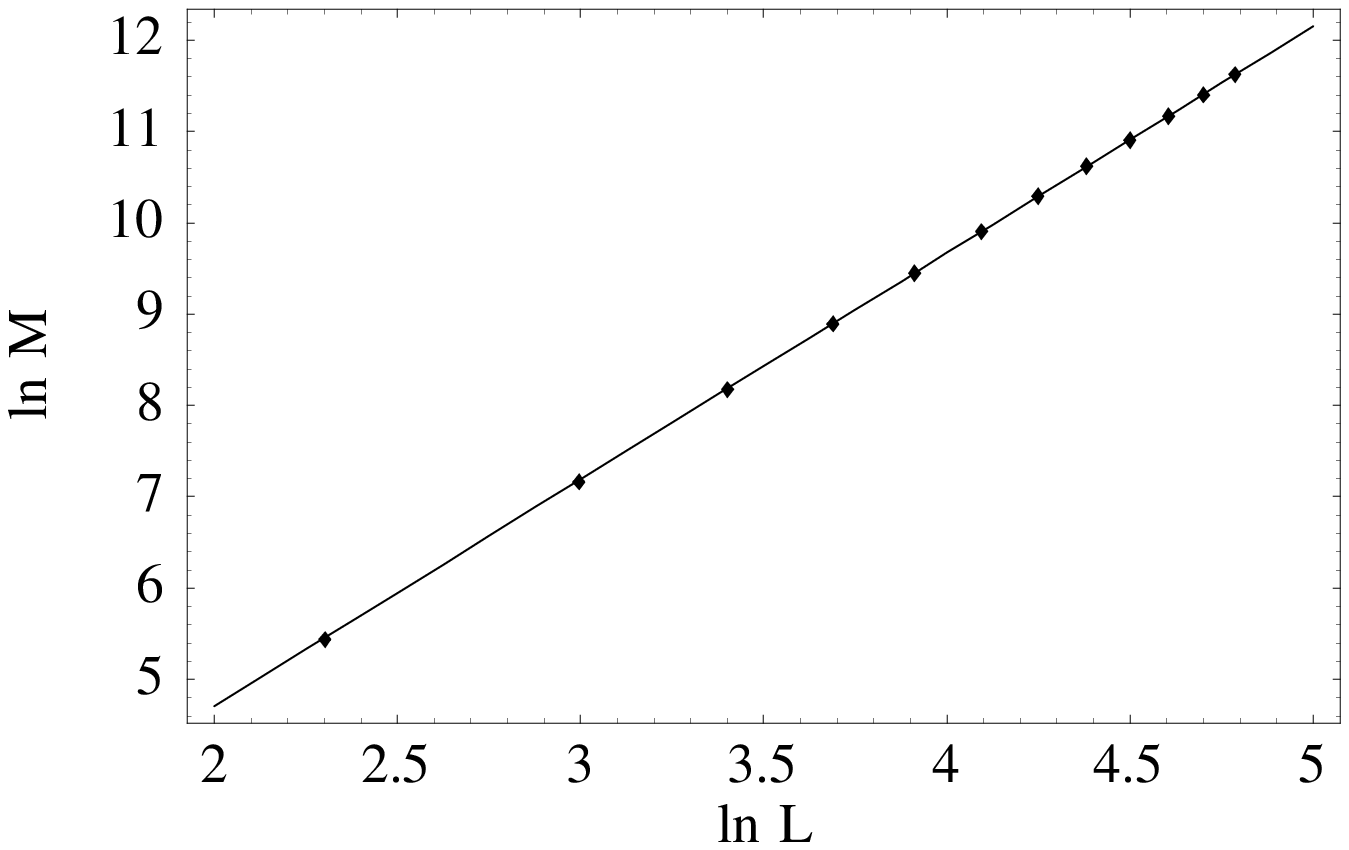}
\caption{Mass of the spanning cluster $\ln M$ vs.\ $\ln L$ for \3D soft spin model.
The solid line is a linear fit.}
\label{fig:MLsoft}
\end{figure}

In our simulations of the soft spin model we used $\lambda=2.0$.
The maximum linear size of the system was $L=120$ and each run consisted of 
10000 MC steps.
Figure~\ref{fig:softK} shows the average critical coupling $\langle \k 
\rangle$ vs. $1/L$. A fit to the function $\langle \k \rangle=K_c/(1+a L^{-p})$  yields $
K_{c}(\lambda=2.0)=0.5100(1)$,
$a=-0.9(4)$ and $p=1.3(1)$.
The smallest size included in the fit is $L=30$ and $\chi^2=10.0$,
$\chi^2/d.o.f.=1.4$ and $Q=0.18$.
This value of $K_{c}$ is in agreement with the value reported by Hasenbusch and
T\"{o}r\"{o}k~\cite{HaTo99} $K_{c}=0.5099049(6)$.
Figure~\ref{fig:MLsoft} shows a log-log plot of $M$ vs.\ $L$.
The resulting value for $\eta$,
$\eta(\lambda=2.0)=0.042(8)$ (with $\chi^2=5.8$, $\chi^2/d.o.f.=0.98$ 
and $Q=0.43$)
is in agreement  with  their result $\eta=0.0381(2)$.
A recent work by Campostrini \emph{et. al.}\cite{CHPRV} yielded $\eta=0.0380(4)$.

\section{Conclusions}
\label{sec:conclusions}
In this paper we have introduced and applied an invaded cluster algorithm
for the two and three-dimensional \XY models and a related 
three-dimensional soft spin model. This work extends the range of validity
of the invaded cluster method to include continuous spins and systems
where the phase transition is of the Kosterlitz-Thouless type.  Our
results for critical temperatures and magnetic exponents are in
reasonable agreement with recent values in the literature.  The invaded
cluster algorithm is very efficient for \XY systems, showing no critical
slowing for estimators of the critical temperature and the magnetization
and yielding highly accurate values of the magnetic exponent with
relatively little computational effort.

\section{Acknowledgments}
The authors would like to thank Nikolay Prokof'ev for useful discussions.
This work was supported in part by NSF grant DMR 9978233.

\newpage
\printfigures

\newpage

\newpage

\newpage

\newpage

\newpage

\newpage

\newpage

\newpage

\newpage

\newpage

\pagebreak

\printtables

\printtables
\begin{center}
\begin{table}
\begin{tabular}{rcrcrcr}
\hline
\multicolumn{1}{c}{L}&\hspace{.1in}
&\multicolumn{1}{c}{$\sigma_{\tilde{\kappa}}$}&\hspace{.1in}
&\multicolumn{1}{c}{$\tilde{\kappa}$}&\hspace{.1in}
&\multicolumn{1}{c}{M} \\
\hline
10 &\hspace{.1in} &0.1079(2)&\hspace{.1in}&0.4730(2)&\hspace{.1in}&236.2(2)\\
20 &\hspace{.1in} &0.0621(1)&\hspace{.1in}&0.46195(7)&\hspace{.1in}&1334(1)\\
30 &\hspace{.1in} &0.04658(8)&\hspace{.1in}&0.45890(5)&\hspace{.1in}&3662(3)\\
40 &\hspace{.1in} &0.03810(7)&\hspace{.1in}&0.45751(3)&\hspace{.1in}&7480(4)\\
50 &\hspace{.1in} &0.03281(6)&\hspace{.1in}&0.45668(2)&\hspace{.1in}&13038(10)\\
60 &\hspace{.1in} &0.02904(5)&\hspace{.1in}&0.45620(2)&\hspace{.1in}&20474(18)\\
70 &\hspace{.1in} &0.02635(4)&\hspace{.1in}&0.45584(2)&\hspace{.1in}&30052(25)\\
80 &\hspace{.1in} &0.02409(4)&\hspace{.1in}&0.45558(1)&\hspace{.1in}&41820(36)\\
90 &\hspace{.1in} &0.02240(4)&\hspace{.1in}&0.45538(1)&\hspace{.1in}&56047(46)\\
100 &\hspace{.1in}&0.02080(4)&\hspace{.1in}&0.45521(1)&\hspace{.1in}&72824(64)\\
110 &\hspace{.1in}&0.01957(3)&\hspace{.1in}&0.45511(1)&\hspace{.1in}&92197(80)\\
120 &\hspace{.1in}&0.01854(3)&\hspace{.1in}&0.455006(9)&\hspace{.1in}&114374(96)\\
\hline\\
\end{tabular}
\caption{Numerical data for the three-dimensional \XY model.}
\label{tab:numer3d}
\end{table}
\end{center}

\newpage

\begin{center}
\begin{table}
\begin{tabular}{rcrcrcr}
\hline
\multicolumn{1}{c}{L}&\hspace{.1in}
&\multicolumn{1}{c}{$\sigma_{\tilde{T}}$}&\hspace{.1in}
&\multicolumn{1}{c}{$\tilde{\kappa}\equiv1/\tilde{T}$}&\hspace{.1in}
&\multicolumn{1}{c}{M} \\
\hline
10 &\hspace{.1in} &0.44(5)&\hspace{.1in}&0.788(2)&\hspace{.1in}&55.5(1)\\
20 &\hspace{.1in} &0.32(5)&\hspace{.1in}&0.891(1)&\hspace{.1in}&200.6(5)\\
40 &\hspace{.1in} &0.25(5)&\hspace{.1in}&0.953(1)&\hspace{.1in}&717(2)\\
80 &\hspace{.1in} &0.22(5)&\hspace{.1in}&0.9936(7)&\hspace{.1in}&2631(6)\\
160 &\hspace{.1in} &0.19(5)&\hspace{.1in}&1.0212(8)&\hspace{.1in}&9543(24)\\
240 &\hspace{.1in} &0.19(5)&\hspace{.1in}&1.0350(8)&\hspace{.1in}&20151(50)\\
320 &\hspace{.1in} &0.18(5)&\hspace{.1in}&1.0410(8)&\hspace{.1in}&34779(86)\\
480 &\hspace{.1in} &0.17(5)&\hspace{.1in}&1.0519(9)&\hspace{.1in}&74083(186)\\
640 &\hspace{.1in} &0.16(5)&\hspace{.1in}&1.0571(7)&\hspace{.1in}&126798(311)\\
800 &\hspace{.1in} &0.17(5)&\hspace{.1in}&1.0605(7)&\hspace{.1in}&192100(490)\\
1000 &\hspace{.1in} &0.16(5)&\hspace{.1in}&1.0643(6)&\hspace{.1in}&293519(756)\\
2000 &\hspace{.1in} &0.15(5)&\hspace{.1in}&1.0742(7)&\hspace{.1in}&1073907(2800)\\
\hline\\
\end{tabular}
\caption{Numerical data for the two-dimensional \XY model.}
\label{tab:numer2d}
\end{table}
\end{center}

\newpage

\begin{center}
\begin{table}
\begin{tabular}{rcrcrcr}
\hline
\multicolumn{1}{c}{L}&\hspace{.1in}
&\multicolumn{1}{c}{$\sigma_{\tilde{\kappa}}$}&\hspace{.1in}
&\multicolumn{1}{c}{$\tilde{\kappa}$}&\hspace{.1in}
&\multicolumn{1}{c}{M} \\
\hline
10 &\hspace{.1in}
&0.151(1)&\hspace{.1in}&0.562(1)&\hspace{.1in}&229(1)\\ 20
&\hspace{.1in} &0.0906(6)&\hspace{.1in}&0.5486(9)&\hspace{.1in}&1288(4)\\
30 &\hspace{.1in}
&0.0652(5)&\hspace{.1in}&0.5160(7)&\hspace{.1in}&3540(10)\\ 40
&\hspace{.1in}
&0.0498(3)&\hspace{.1in}&0.5139(5)&\hspace{.1in}&7270(20)\\ 50
&\hspace{.1in}
&0.0437(3)&\hspace{.1in}&0.5151(4)&\hspace{.1in}&12691(40)\\ 60
&\hspace{.1in}
&0.0382(3)&\hspace{.1in}&0.5123(4)&\hspace{.1in}&19988(70)\\ 70
&\hspace{.1in}
&0.0377(2)&\hspace{.1in}&0.5120(4)&\hspace{.1in}&29404(100)\\ 80
&\hspace{.1in}
&0.0320(2)&\hspace{.1in}&0.5116(3)&\hspace{.1in}&40846(140)\\ 90
&\hspace{.1in}
&0.0320(2)&\hspace{.1in}&0.5115(3)&\hspace{.1in}&54384(200)\\ 100
&\hspace{.1in} 
&0.0300(2)&\hspace{.1in}&0.5113(4)&\hspace{.1in}&70784(240)\\ 110
&\hspace{.1in} 
&0.0282(2)&\hspace{.1in}&0.5111(3)&\hspace{.1in}&89484(300)\\ 120
&\hspace{.1in} 
&0.0266(2)&\hspace{.1in}&0.5112(3)&\hspace{.1in}&111851(400)\\
\hline\\
\end{tabular}
\caption{Numerical data for the \3D soft spin model with $\lambda =2.0$.}
\label{tab:numersoft}
\end{table}
\end{center}

\newpage

\begin{center}
\begin{table}
\begin{tabular}{rcrcr}
\hline
L&\hspace*{0.3in}&$\tau_{\k}$&\hspace*{0.3in}&$\tau_{M}$\\
\hline
10&\hspace*{0.3in}&0.195&\hspace*{0.3in}&0.50\\
20&\hspace*{0.3in}&0.129&\hspace*{0.3in}&0.56\\
30&\hspace*{0.3in}&0.045&\hspace*{0.3in}&0.52\\
40&\hspace*{0.3in}&0.052&\hspace*{0.3in}&0.52\\
50&\hspace*{0.3in}&0.045&\hspace*{0.3in}&0.51\\
60&\hspace*{0.3in}&0.028&\hspace*{0.3in}&0.50\\
70&\hspace*{0.3in}&0.025&\hspace*{0.3in}&0.50\\
80&\hspace*{0.3in}&0.021&\hspace*{0.3in}&0.49\\
90&\hspace*{0.3in}&0.028&\hspace*{0.3in}&0.50\\
100&\hspace*{0.3in}&0.029&\hspace*{0.3in}&0.58\\
110&\hspace*{0.3in}&0.010&\hspace*{0.3in}&0.52\\
120&\hspace*{0.3in}&0.032&\hspace*{0.3in}&0.61\\
\hline\\
\end{tabular}
\caption{Autocorrelation times for critical coupling $\tau_{\k}$ and
magnetization $\tau_{M}$ for
the \3D \XY model.}
\label{tab:tau2d}
\end{table}
\end{center}

\newpage
\begin{center}
\begin{table}
\begin{tabular}{rcrcr}
\hline
L&\hspace*{0.3in}&$\tau_{\k}$&\hspace*{0.3in}&$\tau_{M}$\\
\hline
10&\hspace*{0.3in}&0.273&\hspace*{0.3in}&0.73\\
20&\hspace*{0.3in}&0.144&\hspace*{0.3in}&0.95\\
40&\hspace*{0.3in}&0.107&\hspace*{0.3in}&0.57\\
80&\hspace*{0.3in}&0.077&\hspace*{0.3in}&0.71\\
160&\hspace*{0.3in}&0.095&\hspace*{0.3in}&0.88\\
240&\hspace*{0.3in}&0.061&\hspace*{0.3in}&1.15\\
320&\hspace*{0.3in}&0.073&\hspace*{0.3in}&0.80\\
480&\hspace*{0.3in}&0.097&\hspace*{0.3in}&0.99\\
640&\hspace*{0.3in}&0.071&\hspace*{0.3in}&1.07\\
800&\hspace*{0.3in}&0.082&\hspace*{0.3in}&0.82\\
1000&\hspace*{0.3in}&0.060&\hspace*{0.3in}&0.60\\
2000&\hspace*{0.3in}&0.073&\hspace*{0.3in}&0.64\\
\hline\\
\end{tabular}
\caption{Autocorrelation times for critical coupling $\tau_{\k}$ and
magnetization $\tau_{M}$ for
the \2D \XY model.}
\label{tab:tau3d}
\end{table}
\end{center}
\newpage

\newpage

\begin{center}
\begin{table}
\begin{tabular}{rcrcrcrcr}
\hline
Ref.&\hspace{.1in}&method&\hspace{.1in}&$K_c$ 
(\XY)&\hspace{.1in}&$\eta$\\
\hline
This
work (\XY)&\hspace{.1in}&MC&\hspace{.1in}&0.45412(2)&\hspace{.1in}&0.037(2)&\hspace{.1in}\\
This work (soft spin) &\hspace{.1in}&MC&\hspace{.1in}&-&\hspace{.1in}&0.042(8)\\
\cite{Janke90}&\hspace{.1in}&MC&\hspace{.1in}&0.45408(8)&\hspace{.1in}&0.036(14)\\
\cite{CHPRV}&\hspace{.1in}&MC+HT&\hspace{.1in}&-&\hspace{.1in}&0.0380(4)&\\
\cite{BaFe96}&\hspace{.1in}&MC&\hspace{.1in}&0.454165(4)&\hspace{.1in}&0.042(2)\\
\cite{HaTo99}&\hspace{.1in}&MC&\hspace{.1in}&-&\hspace{.1in}&0.0381(2)\\
\cite{GoHa93}&\hspace{.1in}&MC&\hspace{.1in}&0.45420(2)&\hspace{.1in}&0.024(6)\\
\cite{GuZi}&\hspace{.1in}&FT&\hspace{.1in}&-&\hspace{.1in}&0.038(6)\\
\cite{BuCo}&\hspace{.1in}&HT&\hspace{.1in}&0.45419(3)&\hspace{.1in}&0.039(7)\\
\cite{JaKl}&\hspace{.1in}&FT&\hspace{.1in}&-&\hspace{.1in}&0.0349(8)&\\
\cite{KrLa}&\hspace{.1in}&MC&\hspace{.1in}&-&\hspace{.1in}&0.035(5)\\
\hline\\
\end{tabular}
\caption{A summary of recent estimates of the critical coupling for the
\3D \XY model and the
exponent
$\eta$ for the \3D $O(2)$ universality class.}
\label{tab:3d}
\end{table}
\end{center}

\begin{center}
\begin{table}
\begin{tabular}{rcrcrcr}
\hline
Ref.&\hspace{.1in}&method&\hspace{.1in}&$K_c$&\hspace{.1in}&$\eta$ \\
\hline
This work&\hspace{.1in}&MC&\hspace{.1in}&1.120(1)&\hspace{.1in}&0.251(5)\\
\cite{ZhSc98}&\hspace{.1in}&MC&\hspace{.1in}&1.118&\hspace{.1in}&0.238(4)\\
\cite{KeIr97}&\hspace{.1in}&MC&\hspace{.1in}&1.113(6)&\hspace{.1in}&-\\
\cite{HasPinn97}&\hspace{.1in}&MC&\hspace{.1in}&1.1199(1)&\hspace{.1in}&0.233(3)\\
\cite{Kim96}&\hspace{.1in}&MC&\hspace{.1in}&1.106(5)&\hspace{.1in}&-\\
\cite{Ols95}&\hspace{.1in}&MC&\hspace{.1in}&1.1209(1)&\hspace{.1in}&-\\
\hline\\
\end{tabular}
\caption{A summary of recent estimates of the critical coupling and the
exponent
$\eta$ for \2D \XY model on a simple cubic lattice.
According to the Kosterlitz-Thouless theory $\eta =1/4$.}
\label{tab:2d}
\end{table}
\end{center}


\begin{thebibliography}{10}
\expandafter\ifx\csname bibnamefont\endcsname\relax
  \def\bibnamefont#1{#1}\fi
\expandafter\ifx\csname bibfnamefont\endcsname\relax
  \def\bibfnamefont#1{#1}\fi
\expandafter\ifx\csname url\endcsname\relax
  \def\url#1{\texttt{#1}}\fi
\expandafter\ifx\csname urlprefix\endcsname\relax\def\urlprefix{URL }\fi
\expandafter\ifx\csname bibinfo\endcsname\relax \def\bibinfo#1#2{#2}\fi
\expandafter\ifx\csname eprint\endcsname\relax \def\eprint#1{#1}\fi

\bibitem{KoTh}
\bibinfo{author}{\bibfnamefont{J.~M.} \bibnamefont{Kosterlitz}}
  \bibnamefont{and} \bibinfo{author}{\bibfnamefont{D.~J.}
  \bibnamefont{Thouless}}, \bibinfo{journal}{J. Phys. C}
  \textbf{\bibinfo{volume}{6}}, \bibinfo{pages}{1181} (\bibinfo{year}{1973}).

\bibitem{BuCo}
\bibinfo{author}{\bibfnamefont{P.}~\bibnamefont{Butera}} \bibnamefont{and}
  \bibinfo{author}{\bibfnamefont{M.}~\bibnamefont{Comi}},
  \bibinfo{journal}{Phys. Rev. B} \textbf{\bibinfo{volume}{56}},
  \bibinfo{pages}{8212} (\bibinfo{year}{1997}).

\bibitem{GuZi}
\bibinfo{author}{\bibfnamefont{R.}~\bibnamefont{Guida}} \bibnamefont{and}
  \bibinfo{author}{\bibfnamefont{J.}~\bibnamefont{Zinn-Justin}},
  \bibinfo{journal}{J. Phys. A: Math. Gen.} \textbf{\bibinfo{volume}{31}},
  \bibinfo{pages}{8103} (\bibinfo{year}{1998}).

\bibitem{Janke90}
\bibinfo{author}{\bibfnamefont{W.}~\bibnamefont{Janke}},
  \bibinfo{journal}{Phys. Lett. A} \textbf{\bibinfo{volume}{148}},
  \bibinfo{pages}{306} (\bibinfo{year}{1990}).

\bibitem{GoHa93}
\bibinfo{author}{\bibfnamefont{A.~P.} \bibnamefont{Gottlob}} \bibnamefont{and}
  \bibinfo{author}{\bibfnamefont{M.}~\bibnamefont{Hasenbusch}},
  \bibinfo{journal}{Physica A} \textbf{\bibinfo{volume}{201}},
  \bibinfo{pages}{593} (\bibinfo{year}{1993}).

\bibitem{HaTo99}
\bibinfo{author}{\bibfnamefont{M.}~\bibnamefont{Hasenbusch}} \bibnamefont{and}
  \bibinfo{author}{\bibfnamefont{T.}~\bibnamefont{T{\"{o}}r{\"{o}}k}},
  \bibinfo{journal}{J. Phys. A: Math. Gen.} \textbf{\bibinfo{volume}{32}},
  \bibinfo{pages}{6361} (\bibinfo{year}{1999}).

\bibitem{CHPRV}
\bibinfo{author}{\bibfnamefont{M.}~\bibnamefont{Campostrini}},
  \bibinfo{author}{\bibfnamefont{M.}~\bibnamefont{Hasenbusch}},
  \bibinfo{author}{\bibfnamefont{A.}~\bibnamefont{Pelissetto}},
  \bibinfo{author}{\bibfnamefont{P.}~\bibnamefont{Rossi}}, \bibnamefont{and}
  \bibinfo{author}{\bibfnamefont{E.}~\bibnamefont{Vicari}},
  \emph{\bibinfo{title}{Critical behavior of the three-dimensional {XY}
  universality class}} (\bibinfo{year}{2000}),
  \bibinfo{note}{cond-mat/0010360}.

\bibitem{LiSw96}
\bibinfo{author}{\bibfnamefont{J.~A.} \bibnamefont{Lipa}},
  \bibinfo{author}{\bibfnamefont{D.~R.} \bibnamefont{Swanson}},
  \bibinfo{author}{\bibfnamefont{J.}~\bibnamefont{Nissen}},
  \bibinfo{author}{\bibfnamefont{T.~C.~P.} \bibnamefont{Chui}},
  \bibnamefont{and} \bibinfo{author}{\bibfnamefont{U.~E.}
  \bibnamefont{Israelson}}, \bibinfo{journal}{Phys. Rev. Lett.}
  \textbf{\bibinfo{volume}{76}}, \bibinfo{pages}{944} (\bibinfo{year}{1996}).

\bibitem{GoMu93}
\bibinfo{author}{\bibfnamefont{L.~S.} \bibnamefont{Goldner}},
  \bibinfo{author}{\bibfnamefont{N.}~\bibnamefont{Mulders}}, \bibnamefont{and}
  \bibinfo{author}{\bibfnamefont{G.}~\bibnamefont{Ahlers}},
  \bibinfo{journal}{J. Low Temp. Phys.} \textbf{\bibinfo{volume}{93}},
  \bibinfo{pages}{131} (\bibinfo{year}{1993}).

\bibitem{SwCh92}
\bibinfo{author}{\bibfnamefont{D.~R.} \bibnamefont{Swanson}},
  \bibinfo{author}{\bibfnamefont{T.~C.~P.} \bibnamefont{Chui}},
  \bibnamefont{and} \bibinfo{author}{\bibfnamefont{J.~A.} \bibnamefont{Lipa}},
  \bibinfo{journal}{Phys. Rev. B} \textbf{\bibinfo{volume}{46}},
  \bibinfo{pages}{9043} (\bibinfo{year}{1992}).

\bibitem{Wolff}
\bibinfo{author}{\bibfnamefont{U.}~\bibnamefont{Wolff}},
  \bibinfo{journal}{Phys. Rev. Lett.} \textbf{\bibinfo{volume}{62}},
  \bibinfo{pages}{361} (\bibinfo{year}{1989}).

\bibitem{SwWa}
\bibinfo{author}{\bibfnamefont{R.~H.} \bibnamefont{Swendsen}} \bibnamefont{and}
  \bibinfo{author}{\bibfnamefont{J.-S.} \bibnamefont{Wang}},
  \bibinfo{journal}{Phys. Rev. Lett.} \textbf{\bibinfo{volume}{58}},
  \bibinfo{pages}{86} (\bibinfo{year}{1987}).

\bibitem{NeBa99}
\bibinfo{author}{\bibfnamefont{M.~E.~J.} \bibnamefont{Newman}}
  \bibnamefont{and} \bibinfo{author}{\bibfnamefont{G.~T.}
  \bibnamefont{Barkema}}, \emph{\bibinfo{title}{Monte Carlo Methods in
  Statistical Physics}} (\bibinfo{publisher}{Oxford},
  \bibinfo{address}{Oxford}, \bibinfo{year}{1999}).

\bibitem{KaDo}
\bibinfo{author}{\bibfnamefont{D.}~\bibnamefont{Kandel}} \bibnamefont{and}
  \bibinfo{author}{\bibfnamefont{E.}~\bibnamefont{Domany}},
  \bibinfo{journal}{Phys. Rev. B} \textbf{\bibinfo{volume}{43}},
  \bibinfo{pages}{8539} (\bibinfo{year}{1991}).

\bibitem{ChMa97a}
\bibinfo{author}{\bibfnamefont{L.}~\bibnamefont{Chayes}} \bibnamefont{and}
  \bibinfo{author}{\bibfnamefont{J.}~\bibnamefont{Machta}},
  \bibinfo{journal}{Physica A} \textbf{\bibinfo{volume}{239}},
  \bibinfo{pages}{542} (\bibinfo{year}{1997}).

\bibitem{ChMa98a}
\bibinfo{author}{\bibfnamefont{L.}~\bibnamefont{Chayes}} \bibnamefont{and}
  \bibinfo{author}{\bibfnamefont{J.}~\bibnamefont{Machta}},
  \bibinfo{journal}{Physica A} \textbf{\bibinfo{volume}{254}},
  \bibinfo{pages}{477} (\bibinfo{year}{1998}).

\bibitem{MaCh95}
\bibinfo{author}{\bibfnamefont{J.}~\bibnamefont{Machta}},
  \bibinfo{author}{\bibfnamefont{Y.~S.} \bibnamefont{Choi}},
  \bibinfo{author}{\bibfnamefont{A.}~\bibnamefont{Lucke}},
  \bibinfo{author}{\bibfnamefont{T.}~\bibnamefont{Schweizer}},
  \bibnamefont{and} \bibinfo{author}{\bibfnamefont{L.~V.}
  \bibnamefont{Chayes}}, \bibinfo{journal}{Phys. Rev. Lett.}
  \textbf{\bibinfo{volume}{75}}, \bibinfo{pages}{2792} (\bibinfo{year}{1995}).

\bibitem{MaCh96}
\bibinfo{author}{\bibfnamefont{J.}~\bibnamefont{Machta}},
  \bibinfo{author}{\bibfnamefont{Y.~S.} \bibnamefont{Choi}},
  \bibinfo{author}{\bibfnamefont{A.}~\bibnamefont{Lucke}},
  \bibinfo{author}{\bibfnamefont{T.}~\bibnamefont{Schweizer}},
  \bibnamefont{and} \bibinfo{author}{\bibfnamefont{L.~M.}
  \bibnamefont{Chayes}}, \bibinfo{journal}{Phys. Rev. E}
  \textbf{\bibinfo{volume}{54}}, \bibinfo{pages}{1332} (\bibinfo{year}{1996}).

\bibitem{MoMaCh}
\bibinfo{author}{\bibfnamefont{K.}~\bibnamefont{Moriarty}},
  \bibinfo{author}{\bibfnamefont{J.}~\bibnamefont{Machta}}, \bibnamefont{and}
  \bibinfo{author}{\bibfnamefont{L.~Y.} \bibnamefont{Chayes}},
  \bibinfo{journal}{Phys. Rev. E} \textbf{\bibinfo{volume}{59}},
  \bibinfo{pages}{1425} (\bibinfo{year}{1999}).

\bibitem{LiGl}
\bibinfo{author}{\bibfnamefont{T.~B.} \bibnamefont{Liverpool}}
  \bibnamefont{and} \bibinfo{author}{\bibfnamefont{S.~C.}
  \bibnamefont{Glotzer}}, \bibinfo{journal}{Phys. Rev. E}
  \textbf{\bibinfo{volume}{53}}, \bibinfo{pages}{R4255} (\bibinfo{year}{1996}).

\bibitem{FrCaCo}
\bibinfo{author}{\bibfnamefont{G.}~\bibnamefont{Franzese}},
  \bibinfo{author}{\bibfnamefont{V.}~\bibnamefont{Cataudella}},
  \bibnamefont{and} \bibinfo{author}{\bibfnamefont{C.}~\bibnamefont{A.}},
  \bibinfo{journal}{Phys. Rev. E} \textbf{\bibinfo{volume}{57}},
  \bibinfo{pages}{88} (\bibinfo{year}{1998}).

\bibitem{JoGoMaCh}
\bibinfo{author}{\bibfnamefont{G.}~\bibnamefont{Johnson}},
  \bibinfo{author}{\bibfnamefont{H.}~\bibnamefont{Gould}},
  \bibinfo{author}{\bibfnamefont{J.}~\bibnamefont{Machta}}, \bibnamefont{and}
  \bibinfo{author}{\bibfnamefont{L.~K.} \bibnamefont{Chayes}},
  \bibinfo{journal}{Phys. Rev. Lett.} \textbf{\bibinfo{volume}{79}},
  \bibinfo{pages}{2612} (\bibinfo{year}{1997}).

\bibitem{CoKl}
\bibinfo{author}{\bibfnamefont{A.}~\bibnamefont{Coniglio}} \bibnamefont{and}
  \bibinfo{author}{\bibfnamefont{W.}~\bibnamefont{Klein}}, \bibinfo{journal}{J.
  Phys. A: Math. Gen.} \textbf{\bibinfo{volume}{13}}, \bibinfo{pages}{2775}
  (\bibinfo{year}{1980}).

\bibitem{ACCN}
\bibinfo{author}{\bibfnamefont{M.}~\bibnamefont{Aizenman}},
  \bibinfo{author}{\bibfnamefont{J.~T.} \bibnamefont{Chayes}},
  \bibinfo{author}{\bibfnamefont{L.}~\bibnamefont{Chayes}}, \bibnamefont{and}
  \bibinfo{author}{\bibfnamefont{C.~M.} \bibnamefont{Newman}},
  \bibinfo{journal}{J. Stat. Phys.} \textbf{\bibinfo{volume}{50}},
  \bibinfo{pages}{1} (\bibinfo{year}{1988}).

\bibitem{Chayes98}
\bibinfo{author}{\bibfnamefont{L.}~\bibnamefont{Chayes}},
  \bibinfo{journal}{Comm. Math. Phys.} \textbf{\bibinfo{volume}{197}},
  \bibinfo{pages}{623} (\bibinfo{year}{1998}).

\bibitem{Ko74}
\bibinfo{author}{\bibfnamefont{J.~M.} \bibnamefont{Kosterlitz}},
  \bibinfo{journal}{J. Phys. C} \textbf{\bibinfo{volume}{7}},
  \bibinfo{pages}{1046} (\bibinfo{year}{1974}).

\bibitem{ChMaTaCh}
\bibinfo{author}{\bibfnamefont{Y.~S.} \bibnamefont{Choi}},
  \bibinfo{author}{\bibfnamefont{J.}~\bibnamefont{Machta}},
  \bibinfo{author}{\bibfnamefont{P.}~\bibnamefont{Tamayo}}, \bibnamefont{and}
  \bibinfo{author}{\bibfnamefont{L.}~\bibnamefont{Chayes}},
  \bibinfo{journal}{Int. J. Mod. Phys. C} \textbf{\bibinfo{volume}{10}},
  \bibinfo{pages}{1} (\bibinfo{year}{1999}).

\bibitem{BrTa}
\bibinfo{author}{\bibfnamefont{R.}~\bibnamefont{Brower}} \bibnamefont{and}
  \bibinfo{author}{\bibfnamefont{P.}~\bibnamefont{Tamayo}},
  \bibinfo{journal}{Phys. Rev. Lett.} \textbf{\bibinfo{volume}{62}},
  \bibinfo{pages}{1087} (\bibinfo{year}{1989}).

\bibitem{BaFe96}
\bibinfo{author}{\bibfnamefont{H.~G.} \bibnamefont{Ballesteros}},
  \bibinfo{author}{\bibfnamefont{L.~A.} \bibnamefont{Fen\'{a}ndez}},
  \bibinfo{author}{\bibfnamefont{V.}~\bibnamefont{Mart\'{i}n-Mayor}},
  \bibnamefont{and} \bibinfo{author}{\bibfnamefont{A.}~\bibnamefont{Mu\~{n}oz
  Sudupe}}, \bibinfo{journal}{Phys. Lett. B} \textbf{\bibinfo{volume}{387}},
  \bibinfo{pages}{125} (\bibinfo{year}{1996}).

\bibitem{JaKl}
\bibinfo{author}{\bibfnamefont{F.}~\bibnamefont{Jasch}} \bibnamefont{and}
  \bibinfo{author}{\bibfnamefont{H.}~\bibnamefont{Kleinert}},
  \bibinfo{journal}{J. Math. Phys.} \textbf{\bibinfo{volume}{42}},
  \bibinfo{pages}{52} (\bibinfo{year}{2001}).

\bibitem{KrLa}
\bibinfo{author}{\bibfnamefont{M.}~\bibnamefont{Krech}} \bibnamefont{and}
  \bibinfo{author}{\bibfnamefont{D.~P.} \bibnamefont{Landau}},
  \bibinfo{journal}{Phys. Rev. B} \textbf{\bibinfo{volume}{60}},
  \bibinfo{pages}{3375} (\bibinfo{year}{1999}).

\bibitem{ZhSc98}
\bibinfo{author}{\bibfnamefont{B.}~\bibnamefont{Zheng}},
  \bibinfo{author}{\bibfnamefont{M.}~\bibnamefont{Schultz}}, \bibnamefont{and}
  \bibinfo{author}{\bibfnamefont{S.}~\bibnamefont{Trimper}},
  \bibinfo{journal}{Phys. Rev. E} \textbf{\bibinfo{volume}{59}},
  \bibinfo{pages}{R1351} (\bibinfo{year}{1999}).

\bibitem{KeIr97}
\bibinfo{author}{\bibfnamefont{R.}~\bibnamefont{Kena}} \bibnamefont{and}
  \bibinfo{author}{\bibfnamefont{A.~C.} \bibnamefont{Irving}},
  \bibinfo{journal}{Nucl. Phys. B} \textbf{\bibinfo{volume}{485}},
  \bibinfo{pages}{583} (\bibinfo{year}{1997}).

\bibitem{HasPinn97}
\bibinfo{author}{\bibfnamefont{M.}~\bibnamefont{Hasenbusch}} \bibnamefont{and}
  \bibinfo{author}{\bibfnamefont{K.}~\bibnamefont{Pinn}}, \bibinfo{journal}{J.
  Phys. A: Math. Gen.} \textbf{\bibinfo{volume}{30}}, \bibinfo{pages}{63}
  (\bibinfo{year}{1997}).

\bibitem{Kim96}
\bibinfo{author}{\bibfnamefont{J.~K.} \bibnamefont{Kim}},
  \bibinfo{journal}{Phys. Lett. A} \textbf{\bibinfo{volume}{223}},
  \bibinfo{pages}{261} (\bibinfo{year}{1996}).

\bibitem{Ols95}
\bibinfo{author}{\bibfnamefont{P.}~\bibnamefont{Olsson}},
  \bibinfo{journal}{Phys. Rev. B} \textbf{\bibinfo{volume}{52}},
  \bibinfo{pages}{4526} (\bibinfo{year}{1995}).

\end{thebibliography}
\end{document}